\documentclass[sigconf]{acmart}
\AtBeginDocument{%
  \providecommand\BibTeX{{%
    \normalfont B\kern-0.5em{\scshape i\kern-0.25em b}\kern-0.8em\TeX}}}

\setcopyright{rightsretained}

\copyrightyear{2024}

\acmYear{2024}

\acmDOI{}

\acmConference[CCAI 2024]{CHI 2024 Workshop on Child-centred AI Design}{May 11, 2024}{Honolulu, HI, USA}

\acmBooktitle{CHI 2024 Workshop on Child-centred AI Design, May 11, 2024, Honolulu, HI, USA}
\begin{document}

\title{Inclusive Practices for Child-Centered AI Design and Testing}

\author{Emani Dotch}
\email{dotche@uci.edu}
\orcid{0000-0003-1429-7581}
\affiliation{%
  \institution{University of California, Irvine}
  \country{USA}
}

\author{Vitica Arnold}
\email{vxarnold@uci.edu}
\orcid{0000-0002-6023-6459}
\affiliation{%
  \institution{University of California, Irvine}
  \country{USA}
}
\renewcommand{\shortauthors}{Dotch and Arnold}

\begin{abstract}
   We explore ideas and inclusive practices for designing and testing child-centered artificially intelligent technologies for neurodivergent children. AI is promising for supporting social communication, self-regulation, and sensory processing challenges common for neurodivergent children. The authors, both neurodivergent individuals and related to neurodivergent people, draw from their professional and personal experiences to offer insights on creating AI technologies that are accessible and include input from neurodivergent children. We offer ideas for designing AI technologies for neurodivergent children and considerations for including them in the design process while accounting for their sensory sensitivities. We conclude by emphasizing the importance of adaptable and supportive AI technologies and design processes and call for further conversation to refine child-centered AI design and testing methods.
\end{abstract}




\maketitle
\section{Introduction}
There are a number of ways in which artificial intelligence (AI) can and has been used in child-centered technologies, especially neurodivergent children. Namely, AI has been integrated into technologies to support social communication skills in autistic children \cite{blending_Porayska-Pomsta_2018,mohammed_virtualvoice_2023}. Beyond supporting social communication skills, AI can be useful for neurodivergent children in supporting self-regulation, emotion regulation, and sensory processing needs which are known challenges for this population \cite{cibrian2020supporting, dotch2023supporting,rinaldi_autistic_2022,morgan_connections_2019}. Yet, despite recent advances, child-centered technologies designed with neurodivergence and their specific needs are still limited. Additionally, neurodivergent children are sometimes omitted from the design of child-centered AI technologies, with researchers primarily seeking input from proxies such as caregivers or teachers. There is a variety of work on methods for collaborative design, but we should think about and discuss our methods for designing child-centered AI systems for neurodiverse children in more depth. In our work; we have examined inclusive design that centers neurodivergence in a variety of contexts. In this paper, we describe considerations for designing for neurodivergent children, highlighting those concerns for accessibility and inclusion in the design of AI-enabled technologies and the ways in which design practices must also change to incorporate these perspectives. 

We draw from our experiences conducting participatory design workshops with neurodivergent children, as well as our own lived experiences, to provide considerations for integrating children in AI design and testing. As authors, we both identify as neurodivergent and are related to neurodivergent people. Additionally, we have worked with neurodivergent people in our research and community work. Although the recommendations we set forth are applicable to a wide and diverse population of neurodiverse people, we are particularly concerned in our current work with sensory sensitivities and sensory processing differences. Sensory sensitivities can relate to vision, hearing, olfactory, tactile, vestibular, and proprioceptive senses in which people can be hyposensitive (i.e., a lower threshold for sensory input) or hypersensitive (i.e., a higher threshold for sensory input) \cite{morgan_connections_2019}. 

\subsection{Implications for the Design of Neurodivergent Child-Centered Technologies}
Drawing from our experiences as assistive technology designers and researchers for neurodivergent children~\cite{hayes2010interactive, cibrian2020supporting, dotch2023supporting,escobedo2012mosoco}, we have compiled some ideas that we are considering when designing and testing new AI technologies. Though not exhaustive, these ideas can catalyze discussion on designing for neurodivergent child-centered AI technologies.

As mentioned earlier, sensory processing difficulties are often a shared experience among many neurodivergent individuals, which can influence their interpretation and processing of sensory input. One way we think about leveraging AI to accommodate this is to incorporate systems that model and predict stimulation levels to help balance under or over-stimulation. An AI system designed to anticipate stimulation levels could use predictive behavior modeling to understand individual patterns of stimulation to predict and adapt to a child's sensory processing needs. Features of this could look like adaptive user interfaces that can change colors, contrast, or layout based on the user’s preferences, modifiable sensory outputs like varying levels of visual complexity, or predictive models that understand baseline sensory thresholds and use the data over time to anticipate the child’s sensory needs.

Additionally, to design AI technologies inclusively for neurodivergent children, we believe it's essential to consider their unique attention spans and engagement strategies. Thus, we think about incorporating interactive AI elements that offer novelty and real-time adaptability; we can create experiences that hold these children's interest more effectively. Such AI systems could utilize data on neurodivergent children's engagement patterns to proactively adapt novel stimuli when a child's attention appears to diminish. For instance, these systems might employ dynamic content presentation that alters the way content is delivered based on the child's current engagement level, or it could employ adaptive challenges that offer different levels of difficulty and types of problems that adapt to the child's performance, keeping them challenged but not overwhelmed.

Lastly, neurodivergent children often benefit from environments that provide structure yet flexibility and support that minimizes cognitive load. Integrating adaptive algorithms that tailor the AI’s responses to individual children’s reactions to scaffold these areas can lead to more supportive, effective, and neurodivergent-friendly technologies. AI designed to scaffold cognitive load could include several nuanced features like gentle prompts that remind children of the task at hand, guide them through complex activities, suggest taking breaks to prevent cognitive overload, or structured guidance that offers a clear, consistent routine while allowing for adjustments based on real-time data about the child's performance and state of mind, or memory aids that could help children track tasks, instructions, and goals, supporting their working memory.

As noted above, these ideas are only an entry point for inclusive AI design for neurodivergent children, urging researchers and designers to discuss further what factors should be taken into account when designing with neurodivergent children in mind. In practice, researchers and designers should often include the perspectives of neurodivergent children in their projects to flesh out these design ideas for specific issues and projects further, as discussed in the following section
Accommodations and an overall approach towards inclusivity can broaden the scope of who participates in the design and testing of AI technologies. Highly stimulating environments that might ordinarily be considered to spark creativity or be favorable environments for child designers may contribute to access barriers for neurodivergent children, especially those with sensory sensitivities. Thus, inclusive AI design and testing environments must balance sensory stimulation across environmental/physical factors, social considerations, and design tool choices. Factors such as noise levels, group size, and physical space significantly impact how neurodivergent children with sensory sensitivities may engage in design and testing environments. Strategies such as selecting appropriate workshop locations, minimizing stimulants, and providing quiet rooms or self-regulating tools can enhance inclusivity.

During our projects, we took steps to ensure a sensory-friendly design environment for workshops, yet we still had some shortcomings in this regard. As an example, in one project, we worked with groups of three to four children and chose to host workshops in a conference space without windows to minimize excess light. Additionally, we removed the paintings from the room's walls to minimize stimulants that may become visually overstimulating. While we were prepared to create a sensory-friendly workspace, one point we did not consider was the sensory triggers that our participants might encounter while getting from their vehicles to the workshop room. Because we hosted our design sessions on a university campus, some participants had to walk through campus and/or along a busy road to reach the workshop destination. Following the workshops, some participants shared that it was overstimulating. More specifically, the crowds of students leaving classes, the sound of cars along the streets, and students zipping by on scooters or bikes were overstimulating. With this in mind, in future gatherings, we may consider hosting design sessions away from campus in settings such as a library or other community-centric locations. From this experience, we learned precautions must be taken to ensure participants reach the necessary destinations with minimal sensory triggers. 

Finally, we recommend carefully considering materials and their sensory effects, with options like online platforms (e.g., Miro or Google’s Jamboard) or low-stimulating materials offering viable options. For example, in design sessions we have conducted, we've noticed how tools such as markers can make uncomfortable sounds or have strong scents that can become problematic for some neurodivergent children due to the noise and scent of the markers. Therefore, having alternative options, such as crayons or pencils, allowed our participants to engage in meaningful design without the added sensory stressors. With the aforementioned change in materials, we also suggest that designers think of how the various materials we use may interact with one another. Additionally, having designated spaces for individuals to decompress and providing self-regulating tools can further support integrating children with sensory sensitivities in AI design and testing. Providing self-stimulating tools can help participants self-regulate when needed. Due to the subjective nature of sensory sensitivities, not all contexts or triggers can be predicted and accounted for; thus, having a separate space to wind down may be necessary. For our workshops, we utilized a zen booth down the hall from the workshop space and offered fidget pea pods to our participants when they joined the workshop.

\section{Conclusion}
Our research in AI-enabled child-centered technologies for neurodivergent children indicates that children with sensory differences can and should be involved in the design and testing of AI technologies. However, they require special care, both in terms of the design of the technologies themselves but also of the design processes. Overall, AI design and testing that is adaptable and supportive will best serve neurodivergent children.  With these ideas and reflections in mind, we are interested in further discussing what we should consider when designing with neurodivergent children for AI technologies more broadly with others during the second workshop on child-centered AI.  

\bibliographystyle{ACM-Reference-Format}
\bibliography{Refs}


\begin{thebibliography}{8}

%%% the Web address.

\ifx \showCODEN    \undefined \def \showCODEN     #1{\unskip}     \fi
\ifx \showDOI      \undefined \def \showDOI       #1{#1}\fi
\ifx \showISBNx    \undefined \def \showISBNx     #1{\unskip}     \fi
\ifx \showISBNxiii \undefined \def \showISBNxiii  #1{\unskip}     \fi
\ifx \showISSN     \undefined \def \showISSN      #1{\unskip}     \fi
\ifx \showLCCN     \undefined \def \showLCCN      #1{\unskip}     \fi
\ifx \shownote     \undefined \def \shownote      #1{#1}          \fi
\ifx \showarticletitle \undefined \def \showarticletitle #1{#1}   \fi
\ifx \showURL      \undefined \def \showURL       {\relax}        \fi
% invisible to TeX
\providecommand\bibfield[2]{#2}
\providecommand\bibinfo[2]{#2}
\providecommand\natexlab[1]{#1}
\providecommand\showeprint[2][]{arXiv:#2}

\bibitem[Cibrian et~al\mbox{.}(2020)]%
        {cibrian2020supporting}
\bibfield{author}{\bibinfo{person}{Franceli~L Cibrian}, \bibinfo{person}{Kimberley~D Lakes}, \bibinfo{person}{Arya Tavakoulnia}, \bibinfo{person}{Kayla Guzman}, \bibinfo{person}{Sabrina Schuck}, {and} \bibinfo{person}{Gillian~R Hayes}.} \bibinfo{year}{2020}\natexlab{}.
\newblock \showarticletitle{Supporting self-regulation of children with ADHD using wearables: tensions and design challenges}. In \bibinfo{booktitle}{\emph{Proceedings of the 2020 CHI conference on human factors in computing systems}}. \bibinfo{pages}{1--13}.
\newblock


\bibitem[Dotch et~al\mbox{.}(2023)]%
        {dotch2023supporting}
\bibfield{author}{\bibinfo{person}{Emani Dotch}, \bibinfo{person}{Jialuo Hu}, \bibinfo{person}{Avery Mavrovounioti}, \bibinfo{person}{Weijie Du}, \bibinfo{person}{Jazette Johnson}, \bibinfo{person}{Elizabeth Ankrah}, \bibinfo{person}{Aehong Min}, {and} \bibinfo{person}{Gillian~R Hayes}.} \bibinfo{year}{2023}\natexlab{}.
\newblock \showarticletitle{Supporting Noise Sensitivity and Emotion Regulation with Children}. In \bibinfo{booktitle}{\emph{Proceedings of the 22nd Annual ACM Interaction Design and Children Conference}}. \bibinfo{pages}{522--526}.
\newblock


\bibitem[Escobedo et~al\mbox{.}(2012)]%
        {escobedo2012mosoco}
\bibfield{author}{\bibinfo{person}{Lizbeth Escobedo}, \bibinfo{person}{David~H Nguyen}, \bibinfo{person}{LouAnne Boyd}, \bibinfo{person}{Sen Hirano}, \bibinfo{person}{Alejandro Rangel}, \bibinfo{person}{Daniel Garcia-Rosas}, \bibinfo{person}{Monica Tentori}, {and} \bibinfo{person}{Gillian Hayes}.} \bibinfo{year}{2012}\natexlab{}.
\newblock \showarticletitle{MOSOCO: a mobile assistive tool to support children with autism practicing social skills in real-life situations}. In \bibinfo{booktitle}{\emph{Proceedings of the SIGCHI conference on human factors in computing systems}}. \bibinfo{pages}{2589--2598}.
\newblock


\bibitem[Hayes et~al\mbox{.}(2010)]%
        {hayes2010interactive}
\bibfield{author}{\bibinfo{person}{Gillian~R Hayes}, \bibinfo{person}{Sen Hirano}, \bibinfo{person}{Gabriela Marcu}, \bibinfo{person}{Mohamad Monibi}, \bibinfo{person}{David~H Nguyen}, {and} \bibinfo{person}{Michael Yeganyan}.} \bibinfo{year}{2010}\natexlab{}.
\newblock \showarticletitle{Interactive visual supports for children with autism}.
\newblock \bibinfo{journal}{\emph{Personal and ubiquitous computing}}  \bibinfo{volume}{14} (\bibinfo{year}{2010}), \bibinfo{pages}{663--680}.
\newblock


\bibitem[Mohammed F.~Safi and Mustafa(2023)]%
        {mohammed_virtualvoice_2023}
\bibfield{author}{\bibinfo{person}{Badriya Al~Sadrani Mohammed F.~Safi} {and} \bibinfo{person}{Ashraf Mustafa}.} \bibinfo{year}{2023}\natexlab{}.
\newblock \showarticletitle{Virtual voice assistant applications improved expressive verbal abilities and social interactions in children with autism spectrum disorder: a Single-Subject experimental study}.
\newblock \bibinfo{journal}{\emph{International Journal of Developmental Disabilities}} \bibinfo{volume}{69}, \bibinfo{number}{4} (\bibinfo{year}{2023}), \bibinfo{pages}{555--567}.
\newblock
\urldef\tempurl%
\url{https://doi.org/10.1080/20473869.2021.1977596}
\showDOI{\tempurl}
\showeprint{https://doi.org/10.1080/20473869.2021.1977596}
\newblock
\shownote{PMID: 37346256}.


\bibitem[Morgan({[n.\,d.]})]%
        {morgan_connections_2019}
\bibfield{author}{\bibinfo{person}{Heidi Morgan}.} \bibinfo{year}{[n.\,d.]}\natexlab{}.
\newblock \showarticletitle{Connections Between Sensory Sensitivities in Autism; the Importance of Sensory Friendly Environments for Accessibility and Increased Quality of Life for the Neurodivergent Autistic Minority.}
\newblock  \bibinfo{volume}{13}, \bibinfo{number}{1} (\bibinfo{year}{[n.\,d.]}).
\newblock
\showISSN{2375-7833}
\urldef\tempurl%
\url{https://doi.org/10.15760/mcnair.2019.13.1.11}
\showDOI{\tempurl}


\bibitem[Porayska-Pomsta et~al\mbox{.}(2018)]%
        {blending_Porayska-Pomsta_2018}
\bibfield{author}{\bibinfo{person}{Ka\'{s}ka Porayska-Pomsta}, \bibinfo{person}{Alyssa~M. Alcorn}, \bibinfo{person}{Katerina Avramides}, \bibinfo{person}{Sandra Beale}, \bibinfo{person}{Sara Bernardini}, \bibinfo{person}{Mary~Ellen Foster}, \bibinfo{person}{Christopher Frauenberger}, \bibinfo{person}{Judith Good}, \bibinfo{person}{Karen Guldberg}, \bibinfo{person}{Wendy Keay-Bright}, \bibinfo{person}{Lila Kossyvaki}, \bibinfo{person}{Oliver Lemon}, \bibinfo{person}{Marilena Mademtzi}, \bibinfo{person}{Rachel Menzies}, \bibinfo{person}{Helen Pain}, \bibinfo{person}{Gnanathusharan Rajendran}, \bibinfo{person}{Annalu Waller}, \bibinfo{person}{Sam Wass}, {and} \bibinfo{person}{Tim~J. Smith}.} \bibinfo{year}{2018}\natexlab{}.
\newblock \showarticletitle{Blending Human and Artificial Intelligence to Support Autistic Children’s Social Communication Skills}.
\newblock \bibinfo{journal}{\emph{ACM Trans. Comput.-Hum. Interact.}} \bibinfo{volume}{25}, \bibinfo{number}{6}, Article \bibinfo{articleno}{35} (\bibinfo{date}{dec} \bibinfo{year}{2018}), \bibinfo{numpages}{35}~pages.
\newblock
\showISSN{1073-0516}
\urldef\tempurl%
\url{https://doi.org/10.1145/3271484}
\showDOI{\tempurl}


\bibitem[Rinaldi et~al\mbox{.}({[n.\,d.]})]%
        {rinaldi_autistic_2022}
\bibfield{author}{\bibinfo{person}{L.~J. Rinaldi}, \bibinfo{person}{J. Simner}, \bibinfo{person}{S. Koursarou}, {and} \bibinfo{person}{J. Ward}.} \bibinfo{year}{[n.\,d.]}\natexlab{}.
\newblock \showarticletitle{Autistic traits, emotion regulation, and sensory sensitivities in children and adults with Misophonia}.
\newblock  (\bibinfo{year}{[n.\,d.]}).
\newblock
\showISSN{1573-3432}
\urldef\tempurl%
\url{https://doi.org/10.1007/s10803-022-05623-x}
\showDOI{\tempurl}


\end{thebibliography}
\end{document}